# Indagación llevada a cabo con docentes de primaria en formación sobre temas básicos de Astronomía

**Alejandro Gangui[1,2], María C. Iglesias[2] y Cynthia P. Quinteros[2]**

[1]Instituto de Astronomía y Física del Espacio, CONICET-UBA, Ciudad Universitaria, Buenos Aires, Argentina. E-mail: gangui@df.uba.ar. [2]Centro de Formación e Investigación en la Enseñanza de las Ciencias, FCEyN, Universidad de Buenos Aires, Ciudad Universitaria, Buenos Aires, Argentina.

**Resumen**: Trabajamos en el diagnóstico situacional de los docentes de escuela primaria en formación, con el propósito de desarrollar herramientas didácticas que contribuyan a mejorar su educación formal. Presentamos el material con el que llevamos a cabo una indagación sobre los conocimientos generales en temas básicos de astronomía de los futuros docentes. Se trata de un cuestionario escrito y abierto —aunque también se incluyen algunas preguntas cerradas— sobre un grupo selecto pero representativo de nociones básicas de astronomía. En este trabajo se plantean y discuten los resultados de dos pruebas piloto, suministradas a 30 individuos, y se comentan los cambios realizados para el diseño del instrumento final, que fue implementado en otros 51 alumnos normalistas. Un análisis cualitativo de las respuestas reveló varias concepciones alternativas conocidas en la literatura y otras nuevas. Los resultados muestran una notoria dificultad en la explicación del movimiento de la Luna y de sus fases. Los individuos encuestados también encontraron serias dificultades para explicar un par de elementos astronómicos que integran el lenguaje cotidiano, como ser a qué se llama una estrella fugaz y cuál es la verdadera identidad del lucero. Entre las respuestas ofrecidas para explicar las estaciones del año, se encontró una causalidad singular (que, creemos, no fue aun suficientemente estudiada en la literatura específica): muchos encuestados en lugar de proponer un modelo explicativo —una causa: por ejemplo, la inclinación del eje terrestre— que justificara un determinado fenómeno —el efecto: las diferentes estaciones del año—, hacían uso de otro fenómeno/efecto, en este caso en relación al clima, fenómeno que, como sabemos, posee una fuerte componente astronómica. Presentamos aquí los resultados completos para las pruebas piloto y para la prueba final, y concluimos elaborando algunas conclusiones.

**Palabras clave:** investigación educativa, astronomía, docentes de primaria en formación.

**Title:** Investigation carried out with pre-service elementary teachers on some basic astronomical topics

**Abstract:** We perform a situational diagnosis in topics of astronomy of pre-service elementary teachers in order to try and develop didactic tools that better collaborate with their formal education. In this work we present the instrument we designed to put in evidence some of the most frequently used models on a few basic astronomical notions endowed by these prospective teachers. We work with an open written questionnaire comprising a limited but representative group of basic astronomical notions. We discuss the results of two first pilot tests, provided to 30 individuals, and we comment on the necessary changes applied to the instrument in order to design the final questionnaire, which was then provided to another group of 51 pre-service elementary teachers. A detailed qualitative analysis of the answers revealed many well-known alternative conceptions, and others that seem new. We find that prospective teachers have a hard time in trying to explain the movements of the Moon and its phases. They also meet difficulties to recognize and explain a couple of astronomical elements that make part of our ordinary language, like the origin of a shooting star and the real identity of the "lucero" (a standard way of referring to the planet Venus). Amongst the answers offered to explain the causes of the seasons, we found a singular causality, which we think has not been sufficiently emphasized in the literature so far. Many of the inquired people did not advance an explicative model –a cause: say, the tilt of the Earth's axis– to justify a particular phenomenon –the effect: the seasons on the Earth–, but rather made use of another phenomenon/effect, in the present case related to the climate, in order to explain the seasons. However, as we know, this phenomenon/effect (the climate) has a strong astronomical component. We present here the full results of the first two tests and of the final instrument employed, and we draw some conclusions.

**Introducción**

El enfoque de las ideas previas es conocido en el campo de la didáctica de las ciencias. De manera general, se pueden considerar como aquellas concepciones que tienen los estudiantes acerca del cómo y el por qué las cosas son como son. Ellas responden a una lógica de pensamiento, influenciada por las experiencias realizadas en la vida cotidiana (ya sean éstas de origen sensorial –concepciones espontáneas–, cultural –representaciones culturales– o escolar –concepciones analógicas–). Están caracterizadas por principios conceptuales, epistemológicos y ontológicos, constituyendo verdaderos obstáculos para el aprendizaje. Forman parte de lo que el sujeto ya sabe generando interferencias que dificultan el proceso de construcción científica.

Existe una gran variedad de temas de astronomía para los cuales los futuros docentes de la escuela primaria presentan ideas previas. Esta constatación surge de diversos estudios que indagan sobre concepciones no científicas de los alumnos normalistas sobre, por ejemplo, el origen del ciclo día-noche (Atwood and Atwood, 1995; Vega Navarro, 2001), las estaciones del año (Atwood and Atwood, 1996; Martínez-Sebastià, 2004), el sistema Sol-Tierra-Luna (Vega Navarro, 2007), la explicación de las fases de la Luna (Trundle, et al. 2002), como así también investigaciones acerca de varios otros conceptos de astronomía básica (Camino, 1995; Parker and Heywood, 1998; Trumper, 2003; Pedrochi e Danhoni Neves, 2005).

En estas investigaciones se señala, además, que la recurrencia en los niños de las ideas previas en relación con estos temas se prolonga más allá de la escuela secundaria, afectando también a estudiantes de magisterio y a profesores de la enseñanza primaria en actividad. Las mismas se revelan robustas obstaculizando el proceso de aprendizaje. Todo parece indicar que, si no se trabaja a partir de estas ideas naturales propias de los alumnos, en todos los niveles de la educación formal, el aprendizaje se torna lábil y no significativo.

Nuestro interés por llevar adelante una investigación sobre las ideas previas en temas de astronomía surgió de haber realizado un análisis crítico de los diseños curriculares para la enseñanza primaria. Allí se comprobó la importancia otorgada por los mismos a temas relacionados con los fenómenos astronómicos, en muchas de las jurisdicciones de nuestro país. Nosotros tomaremos como ejemplo a la Ciudad Autónoma de Buenos Aires, por tratarse de la zona representativa en donde nuestra universidad se halla inmersa.

El bloque pedagógico La Tierra y el Universo nuclea los contenidos referidos, entre otros, a ciertos fenómenos astronómicos. La organización de dichos contenidos deja entrever claramente la progresión propuesta a lo largo de ambos ciclos de la enseñanza primaria, como así también sus alcances. Los alumnos pasan del reconocimiento de regularidades y cambios (duración de días y noches en distintas estaciones del año, cambios en el aspecto de la Luna a lo largo del mes, etc.) en el primer ciclo, para comenzar a formarse, a partir de 5° Grado (niños de 10 años), una imagen más estructurada de la Tierra y del Universo, dejando para el último año de la primaria (niños de 12 años) las relaciones más complejas relativas al Sistema Solar.

En un análisis posterior de los lineamientos curriculares, esta vez para la formación docente, encontramos algunos de los temas que presentan muy arraigadas ideas previas, que han sido extensamente estudiadas en la literatura específica. Esto ha llevado a cuestionarnos acerca de la formación –no específica– que presentan los futuros docentes de enseñanza primaria, particularmente en la jurisdicción de la Ciudad de Buenos Aires, y por lo tanto sobre la forma en que los diversos contenidos de astronomía son abordados dentro del aula.

La importancia de estudiar las concepciones de los futuros docentes de primaria (sobre los temas de astronomía que más adelante deberán enseñar a sus alumnos) para lograr una enseñanza de calidad, sumado a la escasez de trabajos realizados sobre este tema en nuestro país, justifican llevar a cabo nuestra investigación. En este trabajo se presenta el cuestionario elaborado con la finalidad de indagar el estado de conocimiento de la población docente del nivel primario en formación. Creemos que es la población más relevante para llevar a cabo un estudio de diagnóstico y posterior desarrollo de herramientas para un aprendizaje significativo. Por ello, el objetivo principal de esta investigación es contribuir al diagnóstico situacional de docentes en formación en relación con algunos temas de astronomía (Iglesias et al., 2007), con la intención de desarrollar herramientas didácticas innovadoras.

**Metodología**

En un trabajo anterior (Gangui et al., 2008) se presentó el resultado de una primera prueba piloto llevada a cabo en una Escuela Normal Superior de la Ciudad Autónoma de Buenos Aires. Se empleó un cuestionario escrito que consta de 12 preguntas que cubren unas ocho temáticas diferentes en astronomía básica (ver cuadro 1). En la primera prueba piloto (Grupo 1: 16 personas encuestadas) sólo se buscaba analizar si las preguntas eran debidamente interpretadas. A partir de ella, y teniendo en cuenta las respuestas de los indagados, se decidió llevar a cabo una serie de cambios, puestos a prueba en dos oportunidades, con futuros docentes de escuela primaria de dos Escuelas Normales diferentes (estos alumnos que respondieron el cuestionario constituyen nuestro Grupo 2: 14 encuestados).

Una vez terminada la fase de prueba, confeccionamos el cuestionario final y lo implementamos en Escuelas Normales diferentes de las anteriores, suministrándolo, como lo habíamos hecho previamente, a grupos de alumnos (51 individuos en total) pertenecientes a la carrera de formación docente primaria. De esta manera, teniendo en cuenta las dos pruebas piloto y la prueba final, completamos un total de 81 encuestados.

**Instrumento piloto**

Pasaremos ahora a detallar los diferentes temas de astronomía que componen el cuestionario (ver cuadro 1). Las primeras dos preguntas indagan sobre las ideas de verticalidad y de la gravedad como una fuerza de atracción hacia el centro de la Tierra (Nussbaum, 1979; Sneider and Ohadi, 1998). La referencia a una historieta en la primera pregunta busca distender y dar una idea de familiaridad al interrogado. En particular, el comentario sobre los "habitantes con la cabeza hacia abajo" expresa una idea común que muchos han oído alguna vez. En la instancia de explicación y/o justificación de la respuesta dada, el interrogado se plantea la situación nuevamente y muestra (ya sea con un breve texto o con un dibujo) su concepto natural de verticalidad. En la segunda pregunta, se trata de hacer que el interrogado se replantee la situación de la acción de la gravitación en la Tierra. A partir de su respuesta podemos verificar si su interpretación de la pregunta 1 era casual o si, por el contrario, se trata de una estructura conceptual estable.

Luego, se pasa a un grupo de tres preguntas que permiten indagar las ideas existentes sobre conceptos astronómicos simples propiamente dichos: el movimiento propio de nuestro planeta, el ciclo día-noche, las diferentes estaciones del año, la órbita que la Tierra describe alrededor del Sol y el grado de inclinación del eje del planeta con respecto a su plano de movimiento (la eclíptica). Todos estos conceptos son muchas veces puestos en relación (en dependencia) y son a veces confundidos a la hora de brindar una explicación científica válida acerca de cómo se producen los fenómenos astronómicos que observamos, ya sea durante el transcurso del día o al completarse un año (Parker and Heywood, 1998). Nuevamente, la pregunta 5 del cuadro 1 pretende analizar el grado de consistencia de la respuesta formulada para la pregunta 3. Como sabemos, el modelo interpretativo que más frecuentemente presentan los alumnos y docentes a la hora de dar respuesta a situaciones referidas a los ciclos estacionales, es de origen analógico. La misma lleva a considerar que la cercanía o lejanía de la Tierra respecto del Sol es causa de las estaciones. La secuenciación de preguntas, pretende resolver la paradoja de la existencia de estaciones alternadas en hemisferios opuestos. Es decir, si la explicación es de tipo analógica, debería ser un modelo útil para ambos hemisferios.

Las preguntas 6 a 8 sirven para conocer las ideas que presentan los futuros docentes sobre los movimientos que posee un cuerpo astronómico que nos es muy familiar, la Luna. La mención explícita a una única cara visible de la Luna (pregunta 6) permite poner a prueba si el encuestado relaciona adecuadamente la combinación de movimientos necesarios para explicar el movimiento real, cuasi mensual, del astro. La pregunta 7 introduce en el razonamiento también al Sol; trata el movimiento de este sistema de tres cuerpos y cómo efectivamente se relacionan fenómenos bien conocidos –los eclipses– con la disposición espacial (y temporal también, por supuesto) de esos cuerpos. Esta pregunta, además, guarda relación con la pregunta 10, sobre las fases lunares. Se ha visto en diversos estudios (por ejemplo, Camino, 1995) que estas fases son muchas veces interpretadas como el resultado de la interposición de la Tierra en el camino de la luz solar que, de otra manera, iluminaría la superficie de la Luna: la llamada teoría del eclipse.

La pregunta 8 fue ex profeso separada de la 6 para que la visión de aquella no incidiese en la posible respuesta de esta última. En efecto, ambas preguntas están muy relacionadas, aunque la 6 tiene algunas reminiscencias culturales mientras que la 8 puede resultar un poco más árida y complicada. Sabemos, sin embargo, que la adecuada respuesta a la 8 implica un acabado conocimiento de lo que se quiere indagar en la 6. Se busca también que, al llegar a la pregunta 8, el encuestado reconsidere su respuesta de la 6 en busca de una reafirmación de lo que piensa. Esto sirve a la investigación, pues permite detectar mejor la verdadera concepción del futuro docente en lo que hace a la relación Tierra-Luna y al movimiento coordinado (subordinado) de esta última.

La pregunta 9 busca explorar las ideas que los docentes en formación tienen sobre los verdaderos movimientos de la Tierra. Tiene conexión directa con las preguntas 3, 4 y 5, planteadas previamente.

Las últimas dos preguntas, la 11 y la 12, sirven para indagar aspectos culturales de la astronomía básica, ya sea en forma de definiciones como a qué se llama comúnmente la estrella polar (Frede, 2006) o qué es realmente una estrella fugaz (pregunta 11). La pregunta 12, al igual que lo hacía la pregunta 1 (con otro estilo, por supuesto), nuevamente recurre a un texto para indagar el conocimiento que el encuestado tiene sobre frases o dichos de uso corriente en el lenguaje común. El poema plantea una situación personal que toma lugar en el comienzo de la noche. La idea de incluir el vocablo lucero dentro del contexto de un poema y no preguntarlo directamente tal como se hace en el caso de las estrellas fugaces en la pregunta anterior, responde a querer guiar al indagado, y permitirle situarse en el momento apropiado en el que generalmente se aprecia el planeta Venus, es decir, al atardecer (aunque sabemos que también puede aparecer al alba). La Luna también podría figurar entre las respuestas aceptables, después de todo, "lucero" indica un astro que se ve en el cielo y que brilla de noche de forma muy intensa. Además, es quizás inapropiado querer arrogarse la "justa" interpretación del significado de un poema. Sin embargo, creemos que el uso común -que es en definitiva el que queremos indagar en los alumnos- se refiere sobre todo al planeta Venus.

Con esta serie de preguntas, se busca testear cuánto y cuán bien la astronomía integra el discurso cotidiano de personas no necesariamente relacionadas con la disciplina, pero que en sus clases luego podrán emplear formas de expresión que involucren aspectos astronómicos.

---

1) En una historieta de Mafalda, ella se encuentra preocupada por nuestra ubicación en el globo terráqueo. Cuando descubre dónde estamos, se desespera al pensar que vivimos cabeza abajo.

¿Es cierto que los habitantes del hemisferio sur estamos con la cabeza hacia abajo? Explique.

2) Suponga que se construye un túnel que atraviesa la Tierra diametralmente, ¿cuál sería el movimiento de una bolita que se deja caer en una de las bocas de dicho túnel? ¿Por qué?

3) ¿Por qué hay diferentes estaciones en el año?

4) Explique el ciclo día-noche, es decir, por qué hay días y noches en la Tierra.

5) ¿Por qué hace más calor en verano que en invierno en el hemisferio sur?

6) ¿Cuál es la razón por la cuál una persona siempre observa la misma cara de la Luna?

7) En la madrugada del jueves 21 de febrero de este año (2008) hubo un eclipse de Luna, ¿qué condiciones deben cumplirse para que esto ocurra?

8) ¿Qué movimientos posee el cuerpo de la Luna? Descríbalos.

9) ¿En qué consisten los movimientos de rotación y traslación terrestres? Descríbalos.

10) El día 12 de mayo de este año (2008) podremos observar la Luna en su cuarto creciente.

a) ¿Cómo se produce esta fase de la Luna?

b) ¿Y una fase de Luna nueva?

11) ¿Qué es una estrella fugaz?

12) En el poema "Una despedida" de Jorge Luis Borges podemos leer la siguiente frase: "La noche había llegado con urgencia. Fuimos hasta la verja en esa gravedad de la sombra que ya el lucero alivia" ¿A qué se refiere Borges con "el lucero"?

---

Cuadro 1.- Cuestionario utilizado en prueba piloto (Grupo 1).

**Análisis de las pruebas piloto**

Durante la primera prueba piloto no se evidenció, respecto de la primera pregunta, referencia alguna a la noción investigada. Esto nos indujo a pensar que quizás el recurso de usar una historieta llevó a que los encuestados no la considerasen una pregunta válida, sino más bien poco seria (por estar enmarcada en una historia cómica). Asimismo, se encontró unanimidad en la respuesta "No es cierto que estamos cabeza abajo", lo cual podría explicarse como una superación del modelo no esférico de la Tierra, que se supone presente en los niños de corta edad. Sin embargo, esto no significa que el conocimiento de la dirección del campo gravitatorio de la Tierra sea una estructura conceptual estable. De hecho, ningún encuestado logró dar siquiera una breve explicación.

Al analizar sus respuestas a la pregunta 2, evidenciamos en algunos casos una imposición de un nuevo movimiento para la Tierra con el cual justificar el movimiento de la bolita que se deja caer. Y esto, claramente, no se relaciona con el concepto de verticalidad investigado. Por ejemplo, entre las respuestas encontramos que "la bolita va y viene por el movimiento que realiza la Tierra". Otras respuestas, no obstante, asumían un movimiento recto de la bolita debido a la gravedad aunque no describían en detalle ese movimiento. Por lo tanto, no puede saberse qué suponen que sucede con la trayectoria de la bolita. En consecuencia no podemos conocer sus modelos de explicación. Algunos alumnos, además, no realizaron una interpretación correcta de la situación propuesta, negando inclusive la posibilidad de que un supuesto túnel pudiera construirse para dejar caer la bolita.

En las preguntas dirigidas a explorar los modelos utilizados para explicar las estaciones del año (preguntas 3 y 5) encontramos respuestas que muestran una forma de causalidad singular. Muchos alumnos, en lugar de proponer un modelo explicativo –esto es, una causa– que justificara un determinado fenómeno –vale decir, el efecto observado–, hacían uso de otro fenómeno/efecto, en este caso en relación al clima, fenómeno que, como bien sabemos, posee una importante componente astronómica (ver cuadro 2).

---

Ejemplo de respuesta a la pregunta 3 (¿Por qué hay diferentes estaciones en el año?). "Por los distintos climas".

Ejemplo de respuesta a la pregunta 5: (¿Por qué hace más calor en verano que en invierno en el hemisferio sur?): "Debe ser por el tipo de clima de este hemisferio".

---

Cuadro 2.- Ejemplos de respuestas encontradas con causalidad singular en relación con las estaciones del año (Grupo 1).

Estas respuestas, que sugieren que el clima es la causa/origen de las diferentes estaciones del año, representan aproximadamente el 10% del total de las encuestas para esta pregunta, como veremos más abajo. Pensamos que situaciones como esta, en las que el indagado confunde causa y efecto, se repiten frecuentemente en poblaciones de futuros maestros, no solamente en temas relacionados con la astronomía. Estamos interesados en indagar más sobre estos obstáculos del aprendizaje y actualmente estamos diseñando herramientas para poder poner a prueba algunas hipótesis que han surgido de este trabajo (Gangui, Iglesias y Quinteros, en preparación).

Algunas de las otras respuestas para estas dos preguntas, como se esperaba, hacían uso de la teoría del alejamiento. Es decir, aplicaban la idea de que la forma (exageradamente) elíptica de la órbita de la Tierra alrededor del Sol es la responsable de que en proximidades del perihelio tengamos el verano, y el invierno corresponda al mayor alejamiento entre la Tierra y el Sol (el afelio). Esta es una explicación que, como fue investigada en reiteradas ocasiones (por ejemplo, Camino, 1995), no logra dar cuenta de las diferentes estaciones en hemisferios opuestos de la Tierra.

En el caso de la pregunta vinculada al ciclo día-noche (pregunta 4), no se evidenciaron inconvenientes para su interpretación, y el modelo más frecuente se corresponde con el científico aceptado hoy en día. Sin embargo, algunos pocos alumnos apelaron a un modelo geocéntrico, no sólo para el ciclo día-noche sino también para las estaciones del año, aunque en este último caso lo adjudicamos a una inadecuada (o poco clara) redacción. También consideramos pertinente notar que, para esta misma pregunta, si bien encontramos respuestas que hacen uso del modelo de la rotación de la Tierra, para algunos alumnos este hecho conduce a que el Sol se aprecia de día y la Luna —exclusivamente— de noche.

En el caso de la pregunta 6, sólo uno de los dieciséis encuestados del Grupo 1 hizo mención a la idea más frecuentemente encontrada en otras investigaciones (Parker and Heywood, 1998): "porque la Luna no gira". En relación con la pregunta 8, algunos encuestados identificaron como "movimientos de la Luna" a la clásica sucesión de imágenes o fotografías correspondientes a las cuatro fases que aparecen comúnmente en los libros de texto. Para la pregunta 10, se obtuvieron sólo dos respuestas y, en ambos casos, éstas resultaron confusas. En general, el tema de la Luna en esta primera prueba piloto, no aportó datos útiles para la investigación. Podemos señalar que muchos encuestados no se "animaron" a dar una explicación sin sentirse seguros de sus respuestas. Y así, se justificaron con no recordar el tema en cuestión. Varios de ellos aseguraron —y hasta se mostraron sorprendidos de— nunca haber notado que la Luna nos muestra siempre la misma cara. En el caso de las preguntas 7 y 9, las respuestas obtenidas fueron más elaboradas y, en general, más acordes al modelo científicamente aceptado.

Para las preguntas 11 y 12, obtuvimos respuestas de lo más variadas: desde "estrellas muertas" o incluso "cometas" que penetran la atmósfera terrestre, como explicación de estrellas fugaces (pregunta 11), hasta toda una variedad de objetos luminosos (astronómicos o no) para la respuesta a la pregunta 12 (por ejemplo: "destello de luz", "reflejo de la Luna", o "el Sol"). Justo es mencionar, sin embargo, que también hubo algunas (pocas) respuestas muy bien encaminadas para la pregunta 11. Estos resultados relacionados con aspectos culturales de la astronomía básica, como era de esperarse, no se vieron modificados en las indagaciones posteriores. Vale decir que sí encontramos muy variadas propuestas para la posible identidad de las estrellas fugaces o del lucero, pero muy pocas de estas coincidieron con las explicaciones que hoy brinda la astronomía.

A partir de los resultados obtenidos en esta primera prueba piloto que acabamos de describir, se decidió llevar a cabo una serie de cambios, puestos a prueba en una segunda oportunidad con un nuevo grupo piloto (Grupo 2).

El primer cambio consistió en una modificación de formato, asignando más espacio por cada pregunta, de modo tal que el encuestado sintiera la necesidad de extenderse un poco más en sus respuestas. Con este cambio se buscó fomentar una reflexión mayor a la hora de ofrecer sus propuestas de explicación. Este objetivo se cumplió exitosamente con este nuevo grupo.

Respecto de las dos primeras preguntas del cuadro 1, consideramos pertinente reformular ambas a fin de facilitar su comprensión y poner en evidencia sus modelos explicativos (ver cuadro 3).

---

1) ¿Es cierto que los habitantes del hemisferio sur estamos con la cabeza hacia abajo? Explique.

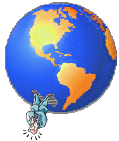

2) Suponga que se construye un pozo que atraviesa la Tierra, ¿cuál sería el recorrido de una bolita que se deja caer desde arriba? Elija una de estas opciones y explique.

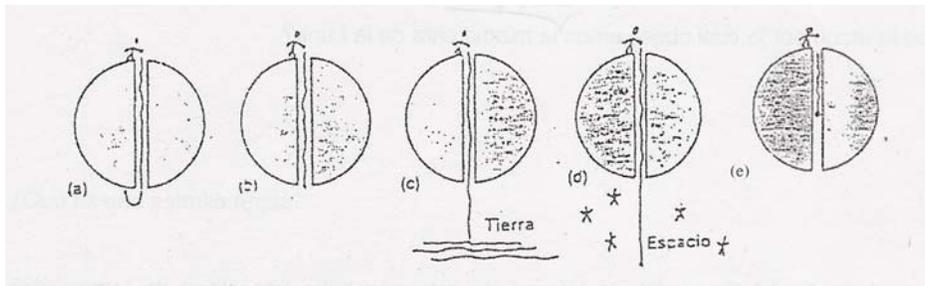

---

Cuadro 3.- Modificaciones al cuestionario para las preguntas 1 y 2 (puestas a prueba con el grupo piloto 2).

Para el caso particular de la pregunta 2, se decidió utilizar la modalidad de opción múltiple debido a las dificultades de interpretación que la misma tuvo para los encuestados. De esta manera, se excluirían, por ejemplo, las respuestas que negaban la posibilidad de la construcción del túnel. El modelo científicamente aceptado explica el movimiento de la bolita como una oscilación entre ambas bocas del túnel en respuesta a la acción de la gravitación como una fuerza dirigida hacia el centro del planeta. Consideramos que los dibujos presentados, al no ser dinámicos, impiden observar exactamente el recorrido que se pretende mostrar en ellos. Por este motivo, decidimos considerar "correctos" aquellos dibujos que más se acercaran a la respuesta científicamente válida, siempre y cuando estuviesen acompañados de una explicación satisfactoria al momento de la indagación (opciones (b) y (e) del cuadro 3). La opción (a) está asociada a una noción de gravedad "superficial" que puede atribuirse a la experiencia cotidiana de que las cosas caen "hacia abajo" (veremos más adelante, en la prueba final, que un diez por ciento de los encuestados elige esta opción).

Es posible considerar que los dibujos proporcionados orientan e inducen la respuesta, aunque sostenemos que la inclusión de varias respuestas posibles les demandaría un tiempo mayor de análisis de la situación problemática propuesta, su justificación y, en consecuencia, nos permitiría obtener más información acerca de sus modelos interpretativos.

En esta segunda experiencia piloto fue posible reconocer una mejora en lo que a la comprensión de los enunciados se refiere. De hecho, hemos encontrado justificaciones más elaboradas al respecto. No obstante, decidimos hacer una modificación adicional en la pregunta 2 para el cuestionario final (ver próxima sección y el cuadro 4).

Merecen especial mención las preguntas dirigidas a explorar las ideas en relación con las fases de la Luna y sus movimientos (preguntas 6, 7, 8 y 10 del cuadro 1). A pesar de lo mencionado para el Grupo 1, no incluimos modificaciones al respecto para esta segunda indagación. Esperábamos quizá una mayor participación en el segundo grupo piloto de futuros docentes. Sin embargo, evidenciamos el mismo inconveniente, muy pocas respuestas, la mayoría incorrectas, y, por lo tanto, se hizo necesaria una revisión en el planteo de estas preguntas (véase la próxima sección).

**Instrumento final**

A partir del análisis y discusión realizados en ambas pruebas piloto, diseñamos el cuestionario final y lo suministramos a dos nuevos grupos de docentes de primaria en formación pertenecientes dos Escuelas Normales Superiores de la Ciudad de Buenos Aires.

En relación con la pregunta 2 del cuadro 3, fue eliminada la opción (c). Esto se justifica en el hecho de que esta concepción resulta ser más extendida entre los niños que entre los adultos (apoyado por el hecho de que ningún encuestado del grupo piloto 2 la había elegido). Asimismo, agregamos una nueva opción (e) en la que se incluye un recuadro para que quien responde pudiera plantear un modelo que considerara adecuado y que no se ajustara a ninguna de las otras opciones propuestas, como también una reformulación del enunciado (ver cuadro 4).

Acerca de las preguntas que pretenden explorar las ideas relacionadas con las fases de la Luna y sus movimientos, consideramos que una instancia de opción múltiple de características similares a las de la pregunta 2 del cuadro 4, al igual que un mayor cuidado en la redacción de las preguntas específicas –con menos lenguaje que dificulte su comprensión (¿a qué llamamos "la cara" de la Luna?)– podría conducir a resultados satisfactorios.

Por este motivo, realizamos modificaciones buscando que el encuestado no evitara responderlas por no haber entendido o comprendido de qué se estaba hablando (ver cuadro 5).

2) Sabemos que esto no es posible, pero lo invitamos a imaginar la siguiente situación hipotética. *"Suponga que se construye un pozo que atraviesa la Tierra, ¿cuál sería el recorrido de una bolita que se deja caer?"* Elija una de estas opciones.

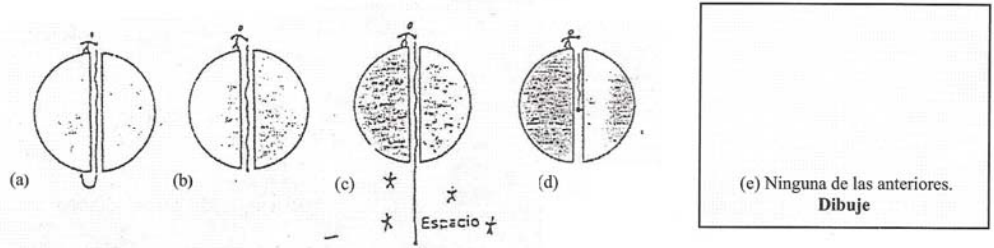

Explique su elección:

Cuadro 4.- Modificaciones al cuestionario para la pregunta 2 (cuestionario final).

Si bien la presencia de imágenes sobre las fases lunares podría configurarse como un elemento distorsionador, su inclusión pretende poner en consideración el fenómeno tal como es observado desde la Tierra, durante un ciclo completo de traslación de la Luna alrededor de nuestro planeta. Es decir, aunque la fase lunar cambie a lo largo de su ciclo, e independientemente de aquello, siempre presenta iluminada la misma "cara". Este es el hecho que se pretende poner en conflicto. La inclusión de una sola imagen de la Luna estaría mostrando solo una parte del problema. El conjunto de imágenes, en cambio, pone a los encuestados en situación de observar, aunque es cierto que con alguna dificultad dada la calidad de las mismas, las partes sombreadas de la Luna a lo largo de su ciclo. O al menos, ofrece la oportunidad de reflexionar sobre ello. De hecho, recién cuando se incluyeron tales imágenes, se lograron respuestas más elaboradas y no relacionadas con las fases lunares.

Respecto de la pregunta que indaga acerca de la razón por la cual observamos siempre la misma cara de la Luna, casi un tercio de los futuros docentes que respondieron el cuestionario final asumieron que se debía a la falta de movimiento de rotación. Por su parte, para la pregunta 10 encontramos que muchos de los indagados sostienen, como era de esperarse, la teoría del eclipse como explicación de las fases de la Luna (esto es, eligieron la última de las opciones consignadas en el cuadro 5).

Mencionemos, por último, que para esta tercera oportunidad, incluimos al pie del cuestionario final un recuadro donde invitábamos a los encuestados a modificar sus respuestas, si ellos lo consideraban apropiado (ver cuadro 6). Si bien fueron encontrados algunos resultados positivos, vale decir respuestas en las que el indagado respetaba este pedido, la mayoría de ellos no lo hizo.

Las demás preguntas del cuestionario no se vieron sustancialmente modificadas en esta oportunidad entendiendo que, a juzgar por lo relevado en las respuestas del grupo piloto 2, se comprendía claramente el objetivo de las mismas.

6) El dibujo representa cuatro fases características de la Luna, que se corresponden con distintas porciones iluminadas de la misma mitad de la Luna (a veces llamada "la cara de la Luna").

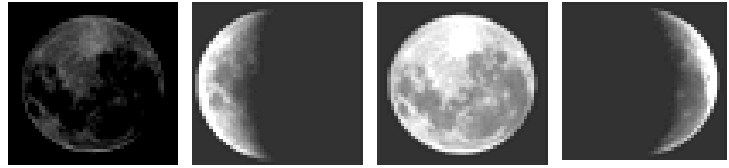

¿Cuál es la razón por la que le observamos siempre la misma cara?

10) ¿A qué se deben las fases lunares? Marque una de las opciones. Justifique su elección con un dibujo.

( ) Las nubes cubren una parte de la Luna y solo vemos la parte iluminada.

( ) El Sol se interpone entre la Tierra y la Luna produciendo sombra.

( ) La Luna refleja la luz del Sol y vemos la zona que está iluminada.

( ) El Sol ilumina la Luna pero la Tierra, al moverse, se interpone produciendo sombra.

Cuadro 5.- Modificaciones al cuestionario referido a las preguntas 6 y 10 que indagan sobre los movimientos de la Luna y sobre sus fases (cuestionario final).

Comentario: llegando al final, en caso que desee modificar algunas de las respuestas que dio, le pedimos que no borre ni tache sino que escriba la nueva respuesta o modificación con otro color.

Cuadro 6.- Recuadro ubicado al pie del cuestionario (cuestionario final).

**Resultados finales**

A continuación se presentan los resultados extraídos a partir del análisis del cuestionario final. Para una mejor interpretación, presentamos los mismos en tablas donde se expresan los porcentajes para cada pregunta, aunque cabe señalar que nuestro estudio no reviste un carácter estadístico (ver anexo 1). Como lo mencionamos más arriba, este cuestionario fue suministrado a dos grupos de docentes de primaria en formación, totalizando 51 nuevos individuos. En los datos que mostraremos abajo hay ciertas preguntas del cuestionario que no cambiaron respecto a las usadas en las pruebas piloto (que, como vimos, incluyeron a 16 y a 14 individuos). Por ello, a veces el número total de datos para una dada pregunta será 51 (en el caso de las preguntas 2, 6, y 8), otras veces será 65 (51+14) y otras veces el número total de datos coincidirá con el de indagados, es decir 81 (preguntas 3, 5, y 7).

*Pregunta que indaga sobre ideas de verticalidad y gravitación (P1 y P2)*

Respecto de la pregunta 1 puede decirse que se ha logrado el objetivo de dejar clara la consigna. A partir de allí las respuestas han sido, en casi un 17%, las aceptadas científicamente. Sin embargo, un 12% de los indagados ha respondido "No" sin ofrecer justificaciones y un 28% de ellos ha respondido "No" pero con justificaciones inadecuadas (ver anexo 1).

Respecto de la pregunta 2, como era de esperarse, muchas explicaciones hicieron referencia a alguna noción particular sobre la gravitación. Por ejemplo, uno de los futuros docentes respondió que la bolita "si atraviesa la tierra caería en otro planeta", sugiriendo la opción (c). Otro escogió la opción (a) y, consignando un argumento que confunde dos interacciones físicas diferentes, ambas presentes en los programas escolares, respondió: "al llegar al otro lado de la tierra, volverá a quedar sobre el suelo como si hubiera un imán".

Podría decirse que estas explicaciones se encuentran en algún estadio entre una concepción primitiva y la concepción científica actual, como fue analizado por ejemplo en (Nussbaum, 1979). Cabe señalar que uno de los encuestados dejó entrever una noción muy próxima a la explicación científica (que, como mencionamos, no estaba entre las opciones del cuestionario), es decir, la oscilación de la bolita alrededor del centro de la Tierra. Eligiendo la opción (d), agregó luego que la bolita "se quedaría en el centro. O quizás llegue al otro lado pero después vuelve al centro".

*Preguntas que permiten indagar las ideas sobre conceptos astronómicos simples (P3, P4 y P5)*

En cuanto a la pregunta: ¿Por qué hay distintas estaciones en el año? (P3) puede decirse que aproximadamente un 41% de las respuestas refieren a la llamada teoría del alejamiento (ver anexo 1). Del mismo modo, en algunos casos consistentemente con lo expresado en la 3, en la pregunta 5: ¿Por qué hace más calor en verano que en invierno en el hemisferio sur? se ha registrado un 36% de respuestas que emplean la misma justificación, esto es, que alegan la lejanía o cercanía del Sol a la Tierra en diferentes épocas del año. Por su parte, en las respuestas a la pregunta ¿Por qué hay días y noches en la Tierra? (P4) un 63% de las respuestas obtenidas corresponde a la explicación científicamente aceptada.

Aunque la frase "en el hemisferio sur" incluida al final de la pregunta 5 podría llevar a los encuestados a una interpretación inadecuada (no buscada a los fines investigativos), entendiendo que se les preguntaba por qué las estaciones de uno y otro hemisferio están invertidas, esta situación no se vio reflejada en las respuestas obtenidas.

*Preguntas que indagan ideas sobre los movimientos que posee la Luna y el sistema de tres cuerpos (P6, P7 y P8).*

Para estas tres preguntas encontramos respuestas similares a las expresadas por los dos grupos piloto (ver anexo 1). Algunos de los futuros docentes (solo un 12% de los encuestados) recuerdan que la Luna posee los movimientos característicos de todos los cuerpos celestes (rotación y traslación), pero no les es simple usar esos conocimientos para predecir o explicar otros fenómenos que surgen a partir de ellos, como el movimiento sincronizado de la Luna alrededor de la Tierra, que hace que siempre nos ofrezca la misma cara. En otras palabras, vemos aquí un ejemplo más que indica que los estudiantes no hacen uso operativo de las hipótesis del modelo para explicar las observaciones conocidas (Martínez-Sebastià, 2004).

Por otra parte, ninguno de los participantes pudo brindar una explicación adecuada y completa a la pregunta de cómo se produce un eclipse de Luna, más allá de sugerir la alineación de los tres astros, que sirve tanto para explicar un eclipse de Luna como uno de Sol. Asimismo, encontramos inconvenientes respecto del vocabulario específico, donde los encuestados confundían los movimientos de rotación y de traslación terrestres. En otros casos, se referían a éstos directamente como "los movimientos circulares" o "el giro de los cuerpos".

*Preguntas que indagan ideas sobre los movimientos de la Tierra (P9) y sobre las fases de la Luna (P10).*

La primera pregunta (P9) pone a prueba los conocimientos del encuestado sobre los movimientos de rotación y traslación de la Tierra. En las dos primeras pruebas piloto que realizamos, esta pregunta obtuvo muy pocas respuestas correctas por parte de los alumnos normalistas (solo un 10% de los 30 indagados respondieron correctamente). Si bien el cambio que efectuamos para el cuestionario final no fue muy radical, pensamos que la forma de hacer la pregunta (y el texto usado) sería relevante, y por ello decidimos separarla en dos preguntas, una sobre la rotación y otra sobre la traslación de la Tierra. Así, en el cuestionario final la P9 se convirtió en: "¿A qué se llama movimiento de rotación terrestre?" (P9a) y "¿A qué se llama movimiento de traslación terrestre?" (P9b). Un cambio simple como este mejoró notoriamente la proporción de respuestas científicas adecuadas, lo que quizá nos dice mucho sobre el gran problema de comprensión de textos –sean científicos o de otras áreas– que existe hoy en la educación secundaria. En nuestros resultados hemos visto que, ya sea por conocer las respuestas de memoria a las nuevas preguntas o bien por su relativa simplicidad, un alto porcentaje de los indagados (casi un 80% de los 51 indagados) respondió correctamente, tanto la P9a como la P9b, empleando ideas científicas adecuadas.

En lo que respecta a la pregunta 10 sobre las fases de la Luna, la instancia de opción múltiple resultó en un mayor número de respuestas correctas (la tercera de la lista de cuatro, ver cuadro 5: "La Luna refleja la luz del Sol y vemos la zona que está iluminada"), aunque pocos justificaron su elección con un dibujo que la avalara. Así lo hicieron 29 del total de 51 indagados, es decir más del 56%, que eligieron la tercera opción. Los pocos esquemas que fueron dibujados resultan, en general, confusos, y algunos sirven incluso para justificar algunas ideas previas, muy comunes en este tema, relacionadas con la teoría del eclipse (ver figura 1).

*Preguntas sobre definiciones y aspectos culturales (P11 y P12).*

Para las preguntas P11 y P12 obtuvimos respuestas que continúan con la tendencia hallada en las pruebas piloto, con los Grupos 1 y 2. Es decir, un espectro extremadamente amplio de respuestas entre las cuales unas pocas se aproximan a la noción mencionada.

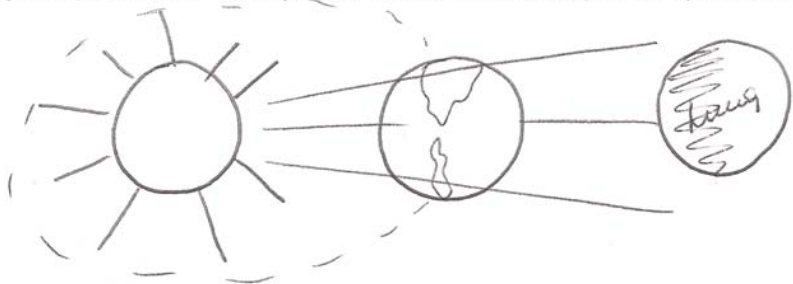

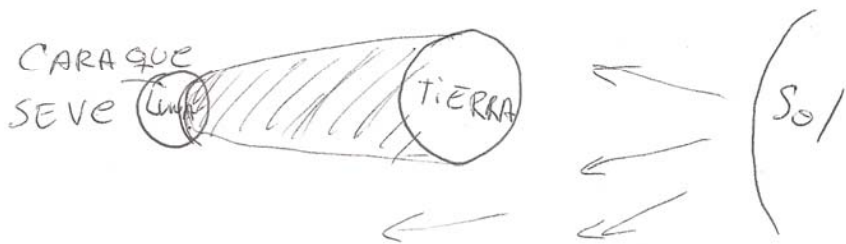

Figura 1.- Dibujos realizados por futuros docentes de escuela primaria para la pregunta que indaga sobre el origen de las fases lunares. El esquema superior incorrectamente emplea la sombra que proyecta la Tierra sobre la superficie de la Luna para justificar el oscurecimiento (eclipse) parcial de esta última, "explicando" las fases lunares (teoría del eclipse). El esquema inferior muestra una situación mixta, donde la opción elegida es la correcta (la tercera de la lista de cuatro opciones), pero el dibujo que la justifica sugiere una idea previa muy arraigada, donde la Tierra nuevamente proyecta una sombra prominente que oscurece parte de la superficie lunar.

La primera pregunta (P11) empleada en el cuestionario final fue expresada con mayor detalle y explicación que la que figura en el cuadro 1. En su versión final quedó así: "En una noche oscura y bien despejada, a veces nos sorprendemos al ver en el cielo una luz que en menos de 1 segundo aparece y desaparece, y que se desplaza con gran rapidez. Son las estrellas fugaces. ¿Qué es una estrella fugaz?" Esta nueva formulación buscaba ayudar al indagado. Sin embargo, no notamos mejoría, pues solo 10 respuestas sobre 51, es decir menos del 20% del total de las respuestas del cuestionario final, sugirieron adecuadamente la posible identidad y origen de una estrella fugaz.

La pregunta 12 sobre el lucero del poema borgeano resultó aun más ardua para los estudiantes normalistas. Si bien existía la posibilidad de que la Luna figurase entre las respuestas aceptables, del total de las respuestas obtenidas, muy pocas hacen mención a muestro satélite natural y casi ningún indagado intentó sugerir en un breve texto la posibilidad de ambigüedad en la identidad del lucero. Solo 3 de ellos, de un total de 51 encuestados, señalaron al planeta Venus como esa luz que alivia la gravedad de la sombra de ese comienzo de anochecer de los arrabales porteños.

**Conclusiones**

Si bien consideramos que la metodología de indagación no es quizás la más adecuada, en tanto no permite acercarse completamente a las formas de pensar y razonar que presentan los docentes al momento de dar una respuesta, creemos que la misma ofrece la oportunidad de contar con un primer diagnóstico descriptivo de sus modelos interpretativos. Asimismo, sirve de sustento para una segunda instancia de investigación. En este sentido, y como indicáramos previamente en este trabajo, estamos abocados a la confección y puesta a prueba de un cuestionario sobre los ciclos estacionales, que se empleará en la modalidad entrevista y que se origina como consecuencia de las indagaciones anteriores. Las estrategias que usaremos involucran entrevistas individuales y voluntarias, que grabaremos en audio y transcribiremos en su totalidad con el fin de poder estudiarlas y extraer aquellas estructuras conceptuales más establecidas. Algo similar estamos implementando actualmente para el caso de la Luna, sus fases y el movimiento del sistema Sol-Tierra-Luna. Consideramos que esta segunda fase permitirá analizar si realmente detrás de estas simples respuestas (escritas y anónimas) existe una representación o modo de pensar y razonar que los explica.

Respecto del trabajo presentado, creemos que los cambios realizados a partir de las dos primeras pruebas piloto fueron positivos ya que observamos, en la tercera oportunidad, mayor participación por parte de los encuestados. Asimismo, hubo una fuerte tendencia a explicar y/o justificar la respuesta elegida mediante discusiones y esquemas que, en algunos casos, favorecieron la reflexión crítica sobre la propia concepción. Esto fue puesto en evidencia por cambios de opinión en las respuestas mientras las formulaban.

A partir de los resultados presentados para el cuestionario final, puede inferirse que muchos de los futuros docentes no poseen (al momento de responder el cuestionario) la formación básica de astronomía que comúnmente se espera que tengan los alumnos de los primeros años de la escuela secundaria. Consideramos que si su formación en estas temáticas no se ve reforzada en los programas de estudio de las Escuelas Normales a las que asisten, un docente de primaria promedio egresado de ellas no dispondrá de herramientas adecuadas para dictar siquiera las nociones más simples de la astronomía. Esto incluye no sólo aspectos observables del cielo, como las fases de la Luna, o la noción de verticalidad y gravitación en el planeta Tierra, sino también la comprensión de ciertos términos y definiciones, procedentes de la astronomía, que ya integran el discurso cotidiano.

Resulta sorprendente, además, que si bien algunos de estos tópicos se encuentran entre los contenidos que deben enseñarse en las escuelas, no esté prevista una adecuada formación y capacitación en los programas de estudio de los profesorados.

Aunque somos conscientes de que los resultados de nuestras indagaciones no representan una muestra significativa o, al menos, representativa para un distrito de las dimensiones de la Ciudad de Buenos Aires, el análisis realizado de los datos obtenidos nos deja entrever que las ideas previas y otras dificultades de aprendizaje están presentes en nuestros futuros docentes y, por lo tanto, merecen especial atención.

Puesto que para los contenidos básicos de la enseñanza primaria los resultados son bastante desalentadores, insistimos en la necesidad de diseñar herramientas de trabajo que permitan subsanar estas deficiencias, como así también dar continuidad a investigaciones en nuestro país que pongan de manifiesto los obstáculos y otras dificultades de enseñanza-aprendizaje en torno a estas cuestiones.

**Agradecimientos**

**Referencias bibliográficas**

**Anexo 1.-** Estadística correspondiente a las preguntas P1, P2, P3, P4, P5, P6, P7 y P8 (incluye pruebas piloto y cuestionario final).

| Pregunta P1 | | |
|---|---|---|
| Respuestas | Absoluto | Porcentaje |
| No | 8 | 12,31 |
| No, pues no existe un arriba o un abajo en el espacio | 11 | 16,92 |
| No (con justificaciones diversas pero inadecuadas) | 18 | 27,69 |
| Sí, pero no nos caemos por la gravedad | 3 | 4,62 |
| Otras | 14 | 21,54 |
| No responde | 11 | 16,92 |
| Total | 65 | 100 |
| **Pregunta P2** | | |
| Respuestas | Absoluto | Porcentaje |
| Opción a) gravedad "superficial" | 5 | 9,80 |
| Opción b) "una forma de gravedad" limitada al planeta | 3 | 5,88 |
| Opción c) gravedad hacia el espacio | 1 | 1,96 |
| Opción d) la fuerza está en el centro | 17 | 33,33 |
| Otras (incluye e) -- tres dibujos) | 16 | 31,37 |
| No responde | 9 | 17,65 |
| Total | 51 | 100 |
| **Pregunta P3** | | |
| Respuestas | Absoluto | Porcentaje |
| Razones climáticas | 8 | 9,88 |
| Cercanía del Sol / traslación de la Tierra | 33 | 40,74 |
| Movimientos de la Tierra | 14 | 17,28 |
| Por la inclinación del eje | 1 | 1,23 |
| Por la combinación de la traslación y la inclinación del eje | 3 | 3,70 |
| Otras | 13 | 16,05 |
| No responde | 9 | 11,11 |
| Total | 81 | 100 |
| **Pregunta P4** | | |
| Respuestas | Absoluto | Porcentaje |
| Por la rotación de la Tierra | 41 | 63,08 |
| Por la Luna | 3 | 4,62 |
| Otras | 18 | 27,69 |
| No responde | 3 | 4,62 |
| Total | 65 | 100 |
| **Pregunta P5** | | |
| Respuestas | Absoluto | Porcentaje |
| Cercanía del Sol | 29 | 35,80 |
| Por la inclinación del eje | 1 | 1,23 |
| Razones climáticas | 6 | 7,41 |
| Por cómo caen los rayos | 15 | 18,52 |
| Por la combinación de la traslación y la inclinación del eje | 0 | 0,00 |
| Otras | 14 | 17,28 |
| No responde | 16 | 19,75 |
| Total | 81 | 100 |

| Pregunta P6 | | |
|---|---|---|
| Respuestas | Absoluto | Porcentaje |
| Porque la Luna no rota | 16 | 31,37 |
| Porque coinciden sus períodos de rotación y traslación | 0 | 0,00 |
| Otras | 26 | 50,98 |
| No responde | 9 | 17,65 |
| Total | 51 | 100 |
| **Pregunta P7** | | |
| Respuestas | Absoluto | Porcentaje |
| Se alinean o interponen la Tierra, la Luna y el Sol | 45 | 55,56 |
| La Luna entra en el cono de sombra de la Tierra | 0 | 0,00 |
| Otras | 14 | 17,28 |
| No responde | 22 | 27,16 |
| Total | 81 | 100 |
| **Pregunta P8** | | |
| Respuestas (en las pruebas piloto)* | Absoluto | Porcentaje |
| Luna nueva, creciente, llena, menguante | 4 | 13,33 |
| Rotación y traslación | 2 | 6,67 |
| Otras | 10 | 33,33 |
| No responde | 14 | 46,67 |
| Total | 30 | 100 |
| | | |
| Respuestas (en el cuestionario final)* | Absoluto | Porcentaje |
| Traslación | 25 | 49,02 |
| Rotación y traslación | 6 | 11,76 |
| Otras | 4 | 7,84 |
| No responde | 16 | 31,37 |
| Total | 51 | 100 |
| | | |
| *Separamos porque se incluyeron fotos en la pregunta 6. | | |